\documentclass{article}
\usepackage{amsmath}
\usepackage{amsfonts}
\usepackage{amssymb}
\def \vs {\vskip5mm}

\def \Sspace {{\cal S}}
\def \be {\begin{equation}}
\def \ee {\end{equation}}
\def \bea {\begin{eqnarray}}
\def \eea {\end{eqnarray}}
\def \mea {\nonumber \\}
\def \ni {\noindent}
\def \chish {{\hat \chi}_{\!{}_S}}
\def \chich {{\hat \chi}_{\!{}_C}}
\def \chis {\chi_{\!{}_S}}
\def \chic {\chi_{\!{}_C}}
\def \AO {{\hat A}}
\def \BO {{\hat B}}
\def \WO {{\hat W}}
\begin{document}
\begin{titlepage}
\title{{\bf Group Theory and  Quasiprobability Integrals of Wigner Functions}}
\date{}
\author{
{\bf Anthony J. Bracken,}$^{\dag *}$
{\bf Demosthenes Ellinas}
\thanks{ellinas@science.tuc.gr;
\dag ajb@maths.uq.edu.au ; \ddag jgw@maths.uq.edu.au}\;
{\bf and James G. Wood}$\ddag$
\\
\\
$\dag\ddag$Centre for Mathematical Physics, Department of Mathematics\\
University of Queensland Brisbane 4072 Australia
\\
\\
$^*$
Department of Sciences, Section of Mathematics \\
Technical University of Crete GR-731 00 Chania Crete Greece}
\maketitle
\begin{center}

{\bf Abstract}
\end{center}

\ni
The integral of the Wigner function of a quantum mechanical system over a region or its boundary
in
the classical phase plane, 
is  called a quasiprobability integral.  Unlike a true probability integral, its value may lie
outside the interval $[0,1]$.
It is characterized by a
corresponding selfadjoint operator, to be called a region or contour operator as appropriate,
which is determined 
by the characteristic function of that region or contour. 
The spectral problem is studied for commuting families of 
region and contour operators associated with concentric disks and circles of given radius $a$.
Their respective eigenvalues are determined as functions of $a$, 
in terms of the Gauss-Laguerre polynomials. 
These polynomials provide a basis of vectors in a Hilbert space 
carrying the positive discrete series representation of 
the algebra ${\mathbf su(1,1)}\approx {\mathbf so(2,1)}$. 
The explicit relation between the spectra of operators associated with
disks and circles with proportional radii,
is given in terms of the discrete variable Meixner polynomials. 


\vs\ni

\end{titlepage}


\ni
\textbf{\bigskip 0. Introduction.}

The problem of bounds on integrals of the Wigner quasidistribution
function \cite{wigner} over some subregion or contour of the
phase plane for 1-dimensional quantum systems, has been
introduced and studied as a theoretical issue \cite{bracken,ellinas}, with
applications in the field of quantum tomography (see {\it e.g.} \cite{leonhardt,sr}).
For a classical (true) probability distribution, such an integral would always have a value in
the interval $[0,1]$, but in the quantum case, the upper and lower bounds vary from region to
region, and may lie outside the interval $[0,1]$.  
The relevance of such studies to quantum tomography,
is that 
these bounds define checks on the accuracy of experimental determinations of Wigner functions.
It has been shown \cite{bracken,ellinas} that, for a given region or contour,  the
bounds are determined by the maximal and minimal eigenvalues of a
corresponding Hermitian operator, to be called the region operator or contour
operator, respectively, acting in
the space of square integrable (wave)functions of the quantum
systems in question. Furthermore, the matrix element of such an
operator \emph{K} between such (wave)functions $\psi$, gives the value of
the non-positive-definite  quasiprobability integral
(QPI), formed by the integral of the Wigner function corresponding to  $\psi$, over
the region or contour that
determines the operator.
Attention is therefore focussed on the spectra of various region and contour operators.  

In what follows, starting from regions and their boundaries in the phase plane of a 
classical mechanical system,
we use their respective characteristic functions to introduce 
corresponding region and contour operators. 
Specifically, we introduce the family of commuting operators $K_{D}(a),\,K_{C}(a)$, 
associated with a disk and a circle respectively,
with radius $a$, 
centered at the
origin of the phase plane.   
These operators are diagonal in
the Fock space of the number states of the quantum harmonic
oscillator.  Their spectra, which determine the upper and lower bounds on the
corresponding QPI, show a remarkable richness of structure.   
We show that these spectra, as
functions of the radius $a$, are determined by 
the Gauss-Laguerre (GL) functions \cite{abramowitz}, acting as basis functions in vector spaces
that carry the $d=1/2$ positive discrete
series representation of the Lie algebra $\textbf{so(2,1)}\approx \textbf{su(1,1)}$, 
involving two different differential operator realizations.
Further, for operators $K_{D,C}(\xi a)$ that refer to disks/circles with 
different radii $\xi a$, $\xi\in\emph{\textbf{R}} $, we find their
eigenvalues are determined by the GL basis functions of scaled
argument $\xi a$, while the latter are expressed as linear combination of the GL
basis of argument $a$, with the Meixner discrete
polynomials as expansion coefficients.

\vs\ni
\textbf{\bigskip 1. Quantum Mechanical setting.}

A one-dimensional quantum system is described by means of the
Heisenberg-Weyl (\textit{HW)} algebra $\mathbf{h_{1}}$, with generators
$\{1,b,b^{\dagger}\}$ satisfying the canonical commutation relations
$bb^{\dagger}-b^{\dagger}b\equiv\lbrack b,b^{\dagger}]=1$,
$[b,1]=0=[b^{\dagger},1]$. In addition to the creation operator
$b^{\dagger}$, annihilation operator  $b$ and unit operator $1,$ we
also use the number operator $N=b^{\dagger}b$, which satisfies the
relations $[N,b]=-b,[N,b^{\dagger}]=$ $b^{\dagger}.$ An irreducible
infinite-dimensional representation $\nu$ of $\ \mathbf{h_{1}}$ is carried by
the so-called Fock space $F\approxeq l_{2}(Z_{+})=\overline{\mathrm{span}%
}\{e_{n}|n\in Z_{+}\}$, with
\begin{align}
\nu(b)e_{n} &  =\sqrt{n}e_{n-1},\;\nu(b^{\dagger})e_{n}=\sqrt{n+1}%
e_{n+1},\;\nu(1)e_{n}=e_{n},\nonumber\\
\nu(N)e_{n} &  =ne_{n},\;\nu(\Pi)e_{n}=(-1)^{n}e_{n},
\end{align}
where we have also introduced the parity operator  $\Pi$
for later use. We adopt the convention to denote an abstract element and its
representative by the same symbol. Then, in the representation $\nu$,
$b^{\dagger}$ is the hermitian conjugate of $b$, \textit{i.e.} $\langle
\psi,b\varphi\rangle_{F}=\langle b^{\dagger}\psi,\varphi\rangle_{F}$, as a
consequence of the orthonormality relations $\langle e_{m},e_{n}\rangle
_{F}=\delta_{mn}$. Now we introduce the bounded displacement
operators
\be
D(\alpha)=\exp(\alpha^{\ast}b-\alpha b^{\dagger}):C\longrightarrow B(F).
\end{equation}
Here $\alpha$ is a complex number and $\alpha^{\ast}$ its complex conjugate.
The displacement operators define a projective representation of the abelian
group of addition in the complex plane:
\begin{align}
D(0) &  =1,\,\,\,D(\alpha)^{\dagger}=D(-\alpha),\\
\text{ \ }D(\alpha)D(\beta) &  =D(\alpha+\beta)e^{i\alpha\times\beta},
\end{align}
with $\alpha,\beta\in C\equiv\Gamma$ and $\alpha\times\beta=\operatorname{Re}%
\alpha\operatorname{Im}\beta-\operatorname{Im}\alpha\operatorname{Re}\beta.$
Note that $\Gamma$ is endowed with the geometric structure of a symplectic
space and physically is identified with the classical phase space of the
quantum mechanical problem dealt with. The family of Wigner operators defined
as $\{\WO(\alpha)=D(\alpha)\Pi D(\alpha)^{\dagger}|\alpha\in\Gamma\}$ provides a
continously parametrized operator basis for the Hilbert space $\Sspace(F)$ of
Hilbert-Schmidt operators on $F$ due to the relation
\be
\langle \WO (\alpha),\WO (\beta)\rangle_{\Sspace}\equiv Tr[\WO (\alpha)^{\dagger
}\WO (\beta)]=\delta(\alpha-\beta)\,.
\end{equation}
Given $\AO, \BO \in\mathcal{S}(F)$, so that $\langle \AO,\AO \rangle_{\Sspace}\,<\infty$
\ \textit{etc.}, then
\be
\AO=\int_{\Gamma}\langle \AO ,\WO(\alpha)\rangle_{\Sspace}\,\WO (\alpha)d^{2}
\alpha\equiv\int_{\Gamma}A(\alpha)\WO (\alpha)d^{2}\alpha
\label{opdef}
\end{equation}
and
\be
\langle \AO,\BO\rangle_{\Sspace}=\int_{\Gamma}A(\alpha)^{\ast}B(\alpha)d^{2}\alpha
\label{scalarproduct}
\end{equation}
where $d^{2}\alpha=d(\operatorname{Re}\alpha)d(\operatorname{Im}
\alpha)/\pi.$ Of special interest are the the density operators $\rho
\in\Sspace(F)$, which comprise the convex set $\mathcal{D}$ of operators
satisfying $\rho^{\dagger}=\rho$, $\rho\,>0$ and $Tr[\rho]=1$. When
expanded in the Wigner operator basis, a density operator furnishes as
expansion coefficient, the corresponding Wigner function\thinspace: 
$W_{\rho}(\alpha)\equiv\langle\rho,\WO (\alpha)\rangle_{\Sspace}.$
Now (\ref{scalarproduct}) gives
\be
{\rm Tr}(\rho \AO )=\int_{\Gamma} A(\alpha)W_{\rho}(\alpha)\,d^2\alpha\,,
\label{average}
\end{equation}
which is interpreted as the expectation value of the observable represented by the operator $\AO$,
or function $A(\alpha)$, when the system is in the state described by the density operator $\rho$, or
by the Wigner function $W_{\rho}$.
The set $P$
of `pure state' density operators have the additional projection operator
property, \textit{i.e} $\rho^{2}=\rho$, and the elements of $P$ are the
extremal points of $\mathcal{D}$, \textit{i.e.} $\mathcal{D}=\mathrm{hull}%
(P).$ For any $\psi\in F$, we define $\psi(x)=\sum_{n}\langle e_{n}%
,\psi\rangle_{F}H_{n}(x)e^{-x^{2}/2}$, where $H_{n}$ is the (normalized)
Hermite polynomial \cite{abramowitz}, so that
\begin{equation}
\langle\psi_{1},\psi_{2}\rangle_{F}=\int_{R}\psi_{1}(x)^{\ast}\psi
_{2}(x)\,dx\,.
\end{equation}
Now if $\rho\in P$ projects onto $\psi\in F$, the Wigner function
corresponding to $\rho$ and $\psi$ takes the form
\begin{equation}
W_{\psi}(q,p)=\frac{1}{2\pi}\int_{R}\psi(q+x/2)^{\ast}\;\psi(q-x/2)
\,e^{ipx}\, dx
\end{equation}
where $q=\operatorname{Re}\alpha,$ $p=\operatorname{Im}\alpha.$ The Wigner
function was introduced as analogous to a probability density in an
exploration of the extent to which quantum mechanics can be cast into the form
of a true statistical theory \cite{wigner,groenewold,moyal}. As is well known
and easy to show, $W_{\rho}(q,p)$ corresponding to some density operator
$\rho$ is not in general positive everywhere. It is consequently called a
quasiprobability function, and some of its previously known bounds, such as
$\ -(1/\pi)\leq W_{\rho}(q,p)\leq (1/\pi)$, as well as some
recently found ones \cite{bracken,ellinas}, 
can be used to help quantify its quasiprobability-density character.

\vs\ni
\textbf{2. Quasiprobability integrals, region operators and contour operators.}
\vs 
Let $\chis (\alpha)$ denote the characteristic function of a given region $S$ in the phase space
$\Gamma$, and let
$\chish$
denote the corresponding operator given by (\ref{opdef}), with $\chis (\alpha)$ replacing $A(\alpha)$
in the RHS.
We refer to
$\chish$
as the  {\it region operator}  corresponding to $S$.  If $S$ is compact,
$\chish$
is bounded, with a discrete spectrum.

Considering (\ref{average}) in the case $\AO=\chish$,  
we see that the 
LHS is the expectation value of $\chish$, in the state $\rho$, while the RHS is the 
QPI on $S$, namely the integral of
$W_{\rho}$ over the support of $\chis$.

It follows that the QPI on $S$ is bounded above and below by the greatest and least
eigenvalues of $\chish$, respectively, and also that these bounds are attained when $\rho$
projects onto the corresponding eigenfunctions of $\chish$.

A region of special interest is the disk of radius $a$, centred on the origin, with region operator
the {\it disk operator} $\chish\equiv K_D (a)$.
It is known \cite{bracken} that $K_D(a)$ commutes with $N$ for every $a$.  This may be viewed as a
consequence of the fact that $N$ generates transformations of $\Gamma$ which leave every such disk
invariant.  Furthermore, the eigenvalue of $K_D(a)$ on the eigenvector $e_n$ of $N$ has been found
to be given by means of the Laguerre   polynomials $L_{n}(x)$ as,
\be
\lambda_{n}^{D}(a)    =2(-1)^{n}\int_{0}^{a}L_{n}(2x^{2})e^{-x^{2}
}x\,dx\,,
\quad n=0,\,1,\,2,\,\dots\,.
\label{discevals}
\end{equation}

Generalising the idea of a region operator,  we define \cite{ellinas}
for every suitably smooth contour $C$ in
$\Gamma$, a  contour operator $\chich$ using (\ref{opdef}), replacing $A(\alpha)$ in
the RHS by the generalised
characteristic function $\chic (\alpha)$ of $C$.
Here $\chic (\alpha)$ is defined by the property that, for every smooth function $F(\alpha)$ on
$\Gamma$,
\be
\int_{\Gamma} F(\alpha)\chic (\alpha)\,d^2\alpha=\int_C F(\alpha)\,dl\,,
\label{contour}
\end{equation}
where $dl$ is the element of length along $C$.
We can write
\be
\chish =\int_S \WO (\alpha)\,d^2\alpha\,,\quad \chich=\int_C \WO (\alpha)\,dl\,.
\label{characops2}
\end{equation}
In the case that $C$ is the circular boundary of the disk of radius $a$, centred on the origin, we can
see that the  circle operator
$\chich\equiv K_C(a)$ is given by
\be
K_C(a)=\lim_{\epsilon\to 0} \frac{1}{\epsilon}\left(K_D(a+\epsilon)-K_D(a)\right)\,.
\label{circleopdef1}
\end{equation}
Furthermore, it can now be seen that $K_C(a)$ also commutes with $N$, and indeed with each $K_D(b)$,
for every value of $a$ and $b$, and that on $\{e_n \}_{n\in Z_{+}}$, $K_C(a)$ has the eigenvalue
\be
\lambda_{n}^{C}(a)    =\frac{d}{da}\lambda_{n}^{D}(a)=2(-1)^{n}L_{n}
(2a^{2})e^{-a^{2}}a\,,
\quad n=0,\,1,\,2,\,\dots\,.
\label{circleevals}
\end{equation}

\vs\ni
\textbf{3. Results.}
\vs
Our main object now
is to  clarify the relation between the spectra  (\ref{discevals})  and (\ref{circleevals})
of
commuting concentric circle and disk operators for various radii.
%
%
%

We begin with some group theoretical preliminaries: Let $\mathbf{g}$ denote
the Lie algebra
$\mathbf{su(1,1)}$ with generators $\{S_{0},S_{\pm}\}$ in some representation, subject to relations
\begin{equation}
\lbrack S_{0},S_{\pm}]=\pm S_{\pm}\;,\;[S_{+},S_{-}]=-2S_{0},\label{su11}%
\end{equation}
together with $\ S_{0}^{\dagger}=S_{0}$ and $S_{\pm}^{\dagger}=S_{\mp}$.
The central element $\mathcal{C}=S_{0}^{2}-\frac{1}{2}S_{+}S_{-}+S_{-}S_{+}$
belongs to the enveloping algebra $U(\mathbf{su(1,1)).}$ Consider in particular
the  positive discrete
series representation $\mathcal{D}^{+}_k$ \ labelled by a positive integer or
half integer $k$, and  carried by a linear space $H$ with formal vectors
$\{d_{n}^{(k)}\}_{n\in Z_{+}}$ and actions of the operators \cite{vilenkin}%
\begin{equation}
S_{\pm}d_{n}^{(k)}=\mu_{\pm}d_{n\pm1}^{(k)},\;S_{0}d_{n}^{(k)}%
=(k+n)d_{n}^{(k)},\;\mathcal{C}d_{n}^{(k)}=k(k-1)d_{n}^{(k)}\,,
\label{seigen}
\end{equation}
with $\mu_{-}=\mu_{-}(n)=[n(2k+n-1)]^{\frac{1}{2}},\mu_{+}=\mu_{-}(n+1).$
We will consider two different realizations of
$\mathcal{D}^{+}_k$, called the $\pi
$-realization and $\sigma$-realization below.  In each, the
generators are realized by appropriate differential operators acting on
representation spaces isomorphic to $l^{2}(Z_{+})$ .
$\bigskip$

$\pi$-{\it Realization}.
For $M$ real nonzero, the central element and the generators of the realization 
$\pi : {\mathbf{g}}\longrightarrow End (H_{\pi})$ 
of $\mathcal{D}^{+}_k$, are denoted $L^{(M)}=\pi(C), L^{(M)}_0=\pi(S_0)$, $L^{(M)}_{\pm}=\pi(S_{\pm})$, 
and act on
the Hilbert space $H_{\pi}\approxeq L_{2}([0,\infty),dr)=\overline{\rm span}\{u_{k,l}%
^{(M)}(r)|M\in R,k\in Z_{+},l\in 
\{-\frac{1}{2},0,\frac{1}{2},1,...\}, r\in(0,\infty)\}$.  The basis vectors
are \cite{vinet}
\begin{equation}
u_{k,l}^{(M)}(r)=(-1)^{k}\left(  \frac{2\sqrt{M}k!}{\Gamma(k+l+\frac{3}{2}%
)}\right)  ^{\frac{1}{2}}(\sqrt{M}r)^{l+1}e^{-\frac{M}{2}r^{2}}L_{k}%
^{l+\frac{1}{2}}(Mr^{2})\,,
\end{equation}
with $L_{a}^{b}(x)$ the associated Laguerre polynomial \cite{abramowitz}.
In this representation
module the generators are realized by differential operators%
\begin{align}
L_{0}^{(M)}  &  =\frac{1}{4M}[-\frac{d^{2}}{dr^{2}}+\frac{l(l+1)}{r^{2}%
}]+\frac{M}{4}r^{2},\\
L_{\pm}^{(M)}  &  =-\frac{1}{4M}[-\frac{d^{2}}{dr^{2}}+\frac{l(l+1)}{r^{2}%
}]+\frac{M}{4}r^{2}\mp\frac{1}{2}(r\frac{d}{dr}+\frac{1}{2}).
\end{align}
Let $d=\frac{1}{2}(l+\frac{3}{2}),$ then the basis vectors of
$H_{\pi}$ are eigenvectors of the central element with eigenvalues
determining the representation {\it i.e.}
\begin{equation}
L^{(M)}u_{k,l}^{(M)}(r)=d(d-1)u_{k,l}^{(M)}(r),
\end{equation}
while the action of the remaining generators on the basis vectors reads%
\begin{equation}
L_{\pm}^{(M)}u_{k,l}^{(M)}(r)=\mu_{\pm}u_{k\pm1,l}^{(M)}(r),\;L_{0}%
^{(M)}u_{k,l}^{(M)}(r)=(d+k)u_{k,l}^{(M)}(r).
\end{equation}

\bigskip

$\sigma$-{\it Realization}.
Here the central element and the generators of the realization
$\sigma : {\mathbf{g}}\longrightarrow End (H_{\sigma})$  of
$\mathcal{D}^{+}_k$ are denoted $J=\sigma(C)$, 
$J_0=\sigma(S_0)$, $J_{\pm}=\sigma(S_{\pm})$, 
and act on
the Hilbert space $H_{\sigma}\approxeq L_{2}%
([0,\infty),r^{w}dr)=\overline{\rm span}\{e_{m}^{(k)}(r)|k\in R_{+},m\in Z_{+},r\in
(0,\infty)\}$.  The  basis vectors are \cite{jeugt}%
\begin{equation}
e_{m}^{(k)}(r)=2^{W}\sqrt{\frac{w\, m!}{\Gamma(2k+m)}}\exp(-r^{w})(2r^{w}%
)^{k-W}L_{m}^{(2k-1)}(2r^{w}),
\end{equation}
with $L_{a}^{b}(x)$ the associated Laguerre polynomial and $W=\frac{w+1}{2w}$
for $w\geq1$. In this representation module the generators are realized by
differential operators
\begin{align}
J_{0}  &  =\frac{1}{2}(w^{-2}r^{2-w}p_{r}^{2}+\xi r^{-w}+r^{w}),\\
J_{1}  &  =\frac{1}{2}(w^{-2}r^{2-w}p_{r}^{2}-\xi r^{-w}-r^{w}),\\
J_{2}  &  =w^{-1}(rp_{r}-\frac{i}{2}(w-1)),
\end{align}
where $p_r = -i(d/dr +1/r)$, $\xi=k(k-1)-W(W-1)$, and the hermitian generators $J_{1}=\frac{1}%
{2}(J_{+}-J_{-})$ \ and $J_{2}=\frac{1}{2i}(J_{+}-J_{-})$ have been introduced.

The basis vectors of $H_{\sigma}$ are eigenvectors of the central
element with eigenvalues determining the representation, {\it i.e.}
\begin{equation}
J e_{n}^{(k)}(r)=k(k-1)e_{n}^{(k)}(r),
\end{equation}
while the action of the rest of generators on the basis vectors reads%
\begin{equation}
J_{\pm}\,e_{n}^{(k)}(r)=\mu_{\pm}\,e_{n\pm
1}^{(k)}(r),\;J_{0}\,e_{n}^{(k)}(r)=(n+k)\,e_{n}^{(k)}(r).
\end{equation}

\bigskip

To make connection with the spectrum of the circle operator $K_{C}(a)$ \ we first
observe that for basis elements of $H_{\pi}$ and $H_{\sigma}$,
the relation $<f(a),g(a)>_{H_{\sigma}}=\newline <af(a),ag(a)>_{H_{\pi}}$ holds among innner products.
The choice of values $w=2$ i.e $W=3/4,$ leads to the relation
$ae_{m}^{(\frac{l}{2}+\frac{3}{4})}(a)=u_{(m,l)}^{(2)}(a)$, among the basis vectors. Then we identify
the eigenvalues of $K_C(a)$  with the basis vectors as follows:
\begin{equation}
\lambda_{n}^{C}(a)=a\sqrt{a}\,e_{n}^{(\frac{1}{2})}(a)=\sqrt{a}\,u_{n,-\frac{1}%
{2}}^{(2)}(a)\,. \label{identification}%
\end{equation}

\vs\ni
{\bf Proposition}
\vs\ni
\textit {Let $\xi\in R_{+}$,  then the spectra $S_{\xi}%
^{C,D}=\{\lambda_{n}^{C,D}(\xi a)|a\in R_{+},n\in Z_{+}\},$ and $S_{1}%
^{C,D}=\{\lambda_{n}^{C,D}(a)|a\in R_{+},n\in Z_{+}\},$ of the respective
circle/disc operators $K_{C,D}(\xi a)$  and $K_{C,D}(a),$  for concentric
circles/disks of radii $\xi a$  and $a,$  are related by means of discrete
Meixner polynomials $\{M_{n}(m,2k;c^{2})|m,n\in Z_{+},k=0,\frac{1}{2},1,...\}$
as follows:%
\begin{equation}
\lambda_{m}^{C,D}(\xi a)=\mathcal{N}_{m}\sum_{n=0}^{\infty}c^{n}%
M_{n}(m,1;c^{2})\lambda_{n}^{C,D}(a)
\end{equation}
where the coefficients $\mathcal{N}_{m},c$ depend on $\xi.$}

\bigskip\textit{Proof}. We start by first stating the following, independent of
any realization of the eigenvalue problem : let $X_{c}=-(c+1/c)S_{0}+S_{+}+S_{-}$ ,
$0<c<1,$ and consider $\mathcal{D}^{(+)}_k$,  the discrete series representation of
$\mathbf{g}$ carried by a space with basis elements $\{d_{n}%
^{(k)}\}_{n\in Z_{+}}$.  
Then 
\begin{equation}
X_{c}v_{m}^{(k)}=(c-1/c)(k+m)v_{m}^{(k)}\,,\quad m\in N,k\in R_{+},\label{xeigen}%
\end{equation}
with
\begin{equation}
v_{m}^{(k)}=\sum_{n=0}^{\infty}\sqrt{\frac{(2k)_{n}}{n!}}c^{n}M_{n}%
(m,2k;c^{2})d_{n}^{(k)},\label{meixner}%
\end{equation}
where $(\alpha)_{n}=\Gamma(\alpha+n)/\Gamma(\alpha)$ is the Pochhammer
symbol \cite{abramowitz}, and the discrete variable Meixner polynomials are defined in terms of
the hypergeometric series \cite{nikiforov}:
\begin{equation}
M_{n}(m,2k;c^{2})=\text{ }_{2}F_{1}[%
\begin{array}
[c]{c}%
-n,-m\\
2k
\end{array}
;1-1/c].
\end{equation}

[The history of the spectral theory of operators like $X_c$ is rich and
interesting in itself,  for it illuminates the relation of special functions
with the representation theory of $\mathbf{su(1,1)}$, and so provides a
mathematical framework for our group theoretical study of region operators.
Our $\pi$-realization is
based on work in \cite{vinet}. In \cite{jeugt2},
the spectral problem of the operator $H=2J_{0}-J_{+}-J_{-}$ \ in the positive
discrete series representation of $\mathbf{su(1,1)}$ was shown to be related to
generalized Laguerre polynomials. Following this, the analogous problem was
addressed \cite{koelink} for $H=\sigma J_{0}-J_{+}-J_{-}$, ($\sigma\in R)$,
and earlier, for a particular representation, in \cite{bacry}. The
explicit relation to Meixner polynomials in some other realizations (cf. our
$\sigma$-realization) was further studied in \cite{jeugt} for the general
operator $H=-(\sigma+1/c)J_{0}-J_{+}-J_{-}$, $(0<\sigma<1)$.]

Since by comparing (\ref{seigen}) and (\ref{xeigen}) we see that operators $S_{0\text{
}}$and $X_{c}$ are isospectral apart from the factor $(c-1/c)$, so we expect
their formal eigenvectors to be related by a unitary transformation. To
determine this tranformation we first use
the Baker-Champell-Hausdorff formula \cite{gilmore}  to rewrite $X_{c}$ in the form%
\begin{equation}
X_{c}=(c-1/c)e^{irS_{2}}S_{0}e^{-irS_{2}},
\end{equation}
where $e^{r}=\frac{1+c}{1-c}.$ Its \ action on the eigenvector $v_{m}^{(k)}$
then yields%
\begin{equation}
S_{0}e^{-irS_{2}}v_{m}^{(k)}=(k+m)e^{-irS_{2}}v_{m}^{(k)},
\end{equation}
which leads to identifying eigenvectors of $S_{0\text{ }}$and $X_{c}$ up to
the  normalization coefficient $\mathcal{N}_{m}$
\begin{equation}
e^{irS_{2}}d_{m}^{(k)}=\mathcal{N}_{m}v_{m}^{(k)}=\mathcal{N}_{m}\sum
_{n=0}^{\infty}<d_{n}^{(k)}\,, e^{irS_{2}}d_{m}^{(k)}>d_{n}^{(k)}. \label{s2}%
\end{equation}
The right hand side of (\ref{s2}) together with (\ref{meixner}) yields%
\begin{equation}
e^{irS_{2}}d_{m}^{(k)}=\phi_{m}\text{ }\mathcal{N}_{m}\sum_{n=0}^{\infty}%
\sqrt{\frac{(2k)_{n}}{n!}}c^{n}M_{n}(m,2k;c^{2})d_{n}^{(k)},
\label{dilation}
\end{equation}
where $|\phi_{m}|=1.$

We first observe that in order to scale the argument of eigenvalues we need
the dilation operator $\exp(a\frac{d}{da})$ and that our choice of $\pi$ and
$\sigma$-realizations are such that their respective operators $L_{2}=\frac
{1}{2}(L_{+}-L_{-})=-\frac{1}{2}(a\frac{d}{da}+\frac{1}{2})$ and $J_{2}%
=\frac{1}{2i}(J_{+}-J_{-})=\frac{1}{2}(a\frac{d}{da}+\frac{3}{2})$ are linear
combination of $a\frac{d}{da},$ hence $\exp(-rL_{2})$ and $\exp(irJ_{2}),$
acting on a function of $a$, take the respective forms%
\begin{equation}
\exp(-rL_{2})f(a)=e^{\frac{r}{4}}f(e^{\frac{r}{2}}a)\text{ and }\exp
(irJ_{2})f(a)=e^{\frac{3r}{4}}f(e^{\frac{r}{2}}a).
\end{equation}

From this equation and the identification (\ref{identification}),
we obtain%
\begin{equation}
\exp(-rL_{2}+\frac{r}{2})\frac{1}{a}u_{n,-\frac{1}{2}}^{(2)}(a)=\exp
(irJ_{2})e_{n}^{(\frac{1}{2})}(a),
\end{equation}
and multiplication of both sides by $a\sqrt{a}$ yields

\begin{equation}
\lambda_{n}^{C}(\xi a)=\xi a\sqrt{\xi a}e_{n}^{(\frac{1}{2})}(\xi a)=\sqrt{\xi
a}u_{n,-\frac{1}{2}}^{(2)}(\xi a).
\end{equation}

Finally, by means of (\ref{s2}) and (\ref{dilation}) and their respective
expressions in $\pi,\sigma$ realizations, and with $\xi=e^{\frac{r}{2}}$
, $c=\frac{e^{r}-1}{e^{r}+1}$ and $\mathcal{N}_{m}=\phi_{m}(1-c^{2})^{\frac
{1}{2}}c^{m},$ where specifically for the $\pi-$realization, $\phi_{m}=(-1)^{m},$
we obtain%

\begin{equation}
\lambda_{m}^{C}(\xi a)=\mathcal{N}_{m}\sum_{n=0}^{\infty}c^{n}M_{n}%
(m,1;c^{2})\lambda_{n}^{C}(a). \label{circlespectrum}%
\end{equation}

From this relation and the spectral decomposition of the circle operator in
the number state basis, we obtain that%

\begin{equation}
K_{C}(e^{\frac{r}{2}}a)    =\sum_{m=0}^{\infty}\lambda_{m}^{C}(e^{\frac{r}%
{2}}a)e_{m}e_{m}^{\dagger}=\sum_{m=0}^{\infty}(\mathcal{N}_{m}\sum
_{n=0}^{\infty}c^{n}M_{n}(m,1;c^{2})\lambda_{n}^{C}(a))e_{m}e_{m}^{\dagger
}\,,
\end{equation}

or that

\begin{equation}
K_{C}(e^{\frac{r}{2}}a)    =\exp(-rL_{2}+\frac{r}{2})K_{C}(a)=\exp
(irJ_{2})K_{C}(a).
\end{equation}

\bigskip In view of the simple relation between circle and disk operator spectra,
we are led to integrate the scaled eigenvalues of
(\ref{circlespectrum}) in order to get analoguous equations for
the eigenvalues of the disk operator, namely
\begin{align}
\lambda_{m}^{D}(\xi a)  &  =\int_{0}^{a}\lambda_{m}^{C}(\xi x)dx=\mathcal{N}%
_{m}\sum_{n=0}^{\infty}c^{n}M_{n}(m,1;c^{2})\int_{0}^{a}\lambda_{n}%
^{C}(x)dx,\mea
&  =\mathcal{N}_{m}\sum_{n=0}^{\infty}c^{n}M_{n}(m,1;c^{2})\lambda_{n}%
^{D}(a)_{\diamond}%
\end{align}

Below we give a diagrammatic form of the content of the above Proposition: 
the two ladders of spectra for the disc and circle 
operators with proportional radii $\lambda_{n}^{C,D}(a)$ and $\lambda_{n}^{C,D}(\xi a)$
respectively, are shown to be related by means of coefficents 
determined by discrete Meixner polynomials, that are 
given as matrix elements of the dilation operator in the 
Gauss-Laguerre basis functions that carry the positive 
discrete series representation of the algebra $\mathbf{su(1,1)}\approx\mathbf{so(2,1)}$.
Also in each spectral ladder the step operators of 
this algebra $J_{\pm}$ or $L_{\pm}$ operate to create and 
annihilate eigenvalues of the respective region or contour operators.

\begin{tabular}
[c]{lllll}%
& $\ \ \vdots$ & $\ \ \ \ \ \ \ \vdots$ & $\ \vdots$ & \\
& $J_{-},L_{-}\downarrow\uparrow J_{+},L_{+}$ &  & $J_{+},L_{+}\uparrow
\downarrow J_{-},L_{-}$ & \\\quad\\
& $\lambda_{2}^{C,D}(\xi a)$ & $\overrightarrow{Meixner\;pol.}$ & $\lambda
_{2}^{C,D}(a)$ & \\\quad\\
& $J_{-},L_{-}\downarrow\uparrow J_{+},L_{+}$ &  & $J_{+},L_{+}\uparrow
\downarrow J_{-},L_{-}$ & \\\quad\\
& $\lambda_{1}^{C,D}(\xi a)$ & $\overrightarrow{Meixner\;pol.}$ & $\lambda
_{1}^{C,D}(a)$ & \\\quad\\
& $J_{-},L_{-}\downarrow\uparrow J_{+},L_{+}$ &  & $J_{+},L_{+}\uparrow
\downarrow J_{-},L_{-}$ & \\\quad\\
& $\lambda_{0}^{C,D}(\xi a)$ & $\overrightarrow{Meixner\;pol.}$ & $\lambda
_{0}^{C,D}(a)$ &
\end{tabular}
\textbf{\ \ \ \ \ \ \ }

\vskip0.4cm

\textbf{\bigskip4. Conclusions.}

 The spectra of operators associated with integrals of Wigner functions over regions and contours 
 of classical phase space, can have interesting structural properties. 
These operators form a new type of quantum mechanical observables,
with important mathematical and physical meanings. 
 In the case of discs and their boundaries, 
their eigenvalues, which carry 
the meaning of the QPI formed by Wigner functions over the corresponding discs and circles,
behave as basis functions of a  representation space of a Lie algebra of generators. 
These generators can be used to step up and down the ladder of these eigenvalues. 
We emphasize this unusual feature: 
the representation is defined on the spectrum of eigenvalues, not on the
underlying function space on which the region or contour operators act.

 Because of their intrinsic interest, as well as their potential importance in quantum
 tomography, it is important to study further the algebraic and functional 
 properties of such operators. 
 A number of open research problems can be mentioned: 
 development of tranformation theory of region and contour operators under maps that induce 
 simple geometric transformations on their associated region or contour of support;
 temporal evolution equations and behaviour of such  operators under 
 Hamiltonian dynamics of their associated quantum systems; 
 clarification of their mathematical relation to covariant phase operators 
 coming from the problem of quantum mechanical angular operators 
 (see e.g \cite{lahti}); addition rules for region and contour operators associated with
 many-body  quantum systems and their relation to geometric manifestations of quantum entanglement; 
 and finally, the extension of the concept of QPI, region and contour operators,
 and their spectral problems
 to phase spaces other than the classical phase plane, such as 
 the sphere and the continuous and discretized torus. 
 We hope to take up some of these problems in future work.

\end{document}